**Systematic VQE Benchmarking of the Deuteron, Triton, and Helium-3 within Lattice Pionless Effective Field Theory**

Pınar Çifci[1], Serkan Akkoyun[1,2,*], Lloyd La Ronde[2]

[1]Department of Physics, Faulty of Science, Sivas Cumhuriyet University, Sivas, Türkiye

[2]School of Mathematics and Physics, University of Surrey, Guildford, Surrey GU2 7XH, United Kingdom

*Corresponding author: sakkoyun@cumhuriyet.edu.tr

**Abstract**

We investigate the performance of quantum algorithms for light nuclear systems by studying the deuteron ($^2H$), triton ($^3H$), and helium-3 ($^3He$) nuclei within a lattice formulation of pionless effective field theory (EFT). We first compute ground-state energies using classical exact diagonalization (ED), serving as a benchmark reference for variational quantum algorithms. We then perform Variational Quantum Eigensolver (VQE) calculations using noiseless classical statevector simulations of quantum circuits, enabling a controlled assessment of algorithmic performance in the absence of hardware-induced noise. We calibrate the two-body low-energy constant using the deuteron system and fit the three-body interaction strength to the triton, then consistently apply the resulting Hamiltonian parameters to the helium-3 nucleus. Our VQE calculations employ physically motivated ansätze targeting the relevant particle-number sector, with explicit particle-number–conserving constructions implemented for the triton and helium-3 systems. The variational optimization includes an analysis of the Hamiltonian energy variance, $\langle H^2 \rangle - \langle H \rangle^2$, providing additional insight into convergence behavior and the quality of the variational states. We find that the VQE results are in good agreement with the corresponding classical ED ground-state energies across all three systems, including the isospin-asymmetric helium-3 nucleus with Coulomb interactions. Overall, our study provides a transparent and reproducible benchmark for assessing the applicability of variational quantum algorithms to few-body nuclear systems. Additionally, we perform a noisy VQE simulation with a depolarizing noise model for the triton system to illustrate the impact of realistic Noisy Intermediate-Scale Quantum (NISQ)-era hardware noise on variational energy estimation.

## 1. Introduction

The application of quantum computing to many-body quantum problems has attracted increasing attention in recent years, motivated by the exponential growth of the Hilbert space with qubit number and the corresponding limitations of classical methods [1]. Variational quantum algorithms, particularly the Variational Quantum Eigensolver (VQE), have been explored for near-term quantum devices, facilitating the approximation of ground-state energies through variational quantum circuits combined with classical optimization [2,3].

In nuclear physics, proof-of-principle investigations have employed quantum simulations for light nuclei—specifically the deuteron and triton—utilizing simplified effective interactions, such as pionless effective field theory (EFT) [4–6]. These works have underscored both the potential and the inherent limitations of noisy intermediate-scale quantum (NISQ) devices, thereby accentuating the necessity for rigorous benchmark studies wherein quantum algorithms are systematically contrasted with exact classical solutions, such as those obtained via exact diagonalization.

Effective field theory (EFT) provides a rigorous framework for such benchmarks. Pionless EFT—a low-energy expansion applicable below the pion mass scale—describes nuclear systems in terms of contact interactions, which allows for a systematic separation between two-body and higher-order forces and systematic calibration of low-energy constants [6,7]. When implemented on a spatial lattice, this framework yields finite-dimensional Hamiltonians that are amenable to both classical exact diagonalization and quantum algorithms, while preserving the key characteristics of few-body nuclear dynamics.

Recent studies employing lattice EFT and quantum Monte Carlo methods have successfully described binding energies and structural properties of light nuclei, such as three- and four-body systems [7–9]. Extending quantum benchmark calculations beyond the deuteron to systems such as triton and helium isotopes introduce additional challenges, including enlarged Hilbert spaces, three-body interactions, isospin asymmetry, and Coulomb effects.

Although the few-body systems studied here remain tractable by classical methods, the primary aim of this work is not to outperform classical solvers but to establish and validate a complete quantum algorithmic workflow — from Hamiltonian construction and qubit mapping to ansatz design, variational optimization, and noise characterization — within a physically meaningful nuclear EFT framework. Such end-to-end benchmarks on classically solvable systems are essential for identifying algorithmic limitations, validating circuit designs, and building the methodological foundation required before extending quantum simulations to classically intractable regimes.

This study presents a systematic benchmark of the variational quantum eigensolver (VQE) applied to deuteron, triton, and helium-3 nuclei using a lattice-regularized pionless effective field theory (EFT) Hamiltonian. Classical exact diagonalization (ED) serves as the reference solver. The two-body interaction is calibrated to the deuteron, and the three-body term is fixed from the triton and applied without modification to helium-3. Through analysis of ground-state energies, variance diagnostics, and selected many-body state properties, we assess the consistency and accuracy of VQE results across these few-body nuclear systems. The present work makes the following contributions: (i) a physics-motivated, particle-number-conserving ansatz construction tailored to each nuclear system's symmetry sector; (ii) a hierarchical LEC calibration protocol in which parameters fitted to lighter systems are transferred without modification to heavier ones, enabling genuine predictive tests; (iii) variance diagnostics as a convergence and eigenstate-quality indicator alongside energy benchmarks; and (iv) a depolarizing noise analysis for the triton system, providing a quantitative assessment of NISQ-era hardware impacts within the EFT framework.

## 2. Theoretical Framework

### 2.1 Low-energy description and lattice setup

At momentum scales much smaller than the inverse range of the nuclear interaction, few-body nuclear systems can be described within an effective field theory framework based on short-range contact interactions. In this regime, the leading-order dynamics is governed by two-body contact interactions, while a three-body contact term is required to obtain renormalization-group invariant descriptions in the three-body sector. The corresponding low-energy constants are fixed by matching to selected low-energy observables, after which additional few-body properties can be explored within the EFT framework [10-12].

We employ a lattice formulation of pionless effective field theory based on established formulations [5,10]. The lattice discretization yields a finite-dimensional Hilbert space and a Hamiltonian representation that is well suited for both classical exact diagonalization and variational quantum simulations. Within this setting, the framework is employed as a controlled model space for benchmarking quantum algorithms against exact classical reference calculations.

In pionless EFT, the low-energy constants C and D are not universal quantities but depend on the regularization scheme and the momentum cutoff scale employed [13,14]. The lattice spacing $a$ and the corresponding ultraviolet momentum cutoff $\Lambda \sim \pi/a$ therefore play a central role in ensuring the theoretical consistency of the framework [15]. For renormalization-group (RG) invariance to be maintained, the cutoff must be chosen at or above the breakdown scale of the EFT [14]. The breakdown scale of pionless EFT is set by the pion mass, with neutral and charged pion masses of approximately 135 MeV and 140 MeV, respectively, defining a natural upper limit of around 135–140 MeV for reliable application of the theory. In the present work, we adopt a lattice spacing of $a = 4.5$ fm, corresponding to a momentum cutoff of $\Lambda \approx 138$MeV consistent with the pion mass scale and therefore within the validity range of pionless EFT.

### 2.2 Lattice EFT Hamiltonian

We consider one-dimensional spatial lattice with lattice spacing $a$. Each lattice site hosts four spin–isospin modes $\sigma \in \{p\uparrow, p\downarrow, n\uparrow, n\downarrow\}$. Fermionic creation and annihilation operators $a_\sigma^\dagger(i)$ and $a_\sigma(i)$ act on mode $\sigma$ at lattice site $i$, and the corresponding number operator is defined as

$$n_\sigma(i) = a_\sigma^\dagger(i)\, a_\sigma(i). \tag{1}$$

The lattice Hamiltonian within the pionless effective field theory (EFT) framework is given by [5,10].

$$\begin{aligned} H_{\mathrm{EFT}} = -t \sum_{\langle i,j\rangle,\sigma} \left[\, a_\sigma^\dagger(i) a_\sigma(j) + a_\sigma^\dagger(j) a_\sigma(i)\right] + \frac{C}{2} \sum_{i,\sigma\neq\sigma'} n_\sigma(i) n_{\sigma'}(i) \\ + \frac{D}{6} \sum_{i,\sigma\neq\sigma'\neq\sigma''} n_\sigma(i) n_{\sigma'}(i) n_{\sigma''}(i), \end{aligned} \tag{2}$$

where $t = 1/(2Ma^2)$ denotes the hopping amplitude, with M representing the nucleon mass and a the lattice spacing. The first term describes the kinetic energy via nearest-neighbor hopping, where ⟨i,j⟩ denotes pairs of adjacent lattice sites. The second term describes an on-site two-body contact interaction with coupling strength C, and the third term introduces an on-site three-body contact interaction with coupling strength D.

For three-body systems with short-range interactions, the inclusion of the three-body force is required to obtain cutoff-independent results [5]. In this work, the parameter D was determined by fitting to the triton binding energy. This value of D was then applied without modification to other three-body systems.

### 2.3 Coulomb interaction and calibration strategy

In the ³He system, the repulsive Coulomb interaction between protons is included as an additional two-body contribution to the lattice Hamiltonian. Derived from the standard second-quantized form of a generic two-body interaction, the Coulomb term can be written in coordinate space as

$$\hat{H}_{\mathrm{C}} = \frac{1}{2}\int d^3\mathrm{x}\, d^3\mathrm{y}\; \hat{\psi}_p^\dagger(\mathrm{x})\, \hat{\psi}_p^\dagger(\mathrm{y})\, \frac{e^2}{|\mathrm{x}-\mathrm{y}|}\, \hat{\psi}_p(\mathrm{y})\, \hat{\psi}_p(\mathrm{x}), \tag{3}$$

This is the standard second-quantized representation of the Coulomb interaction between charged fermions [16].

Upon discretization on a spatial lattice, the proton field operators are mapped onto lattice sites, such that the Coulomb interaction becomes an interaction between proton number operators,

$$H_{Coulomb} = \textstyle\sum_{i<j} V_{ij} n_p(i) n_p(j), \quad n_p(i) = \textstyle\sum_{\sigma\in\{p\uparrow,p\downarrow\}} n_\sigma(i), \tag{4}$$

where $i, j$ label lattice sites and $V_{ij}$ depends on the site separation.

On a discretized lattice, the Coulomb interaction at zero separation cannot be treated directly and therefore necessitates a short-distance regularization. Accordingly, upon co-occupation of a lattice site by two protons, the Coulomb contribution is modeled as a finite on-site term, in accordance with established lattice formulations of the Coulomb potential and its regularization at the origin [17].

In practice, the coefficients are evaluated for on-site, nearest-neighbor, and next-nearest-neighbor separations within the three-site one-dimensional chain. The resulting Coulomb interaction term was then incorporated into the system Hamiltonian employed for both classical exact diagonalization and variational quantum eigensolver simulations.

The low-energy constants are calibrated hierarchically: the two-body coupling C is fixed by reproducing the experimental deuteron binding energy, while the three-body coupling D is determined from the triton binding energy. The resulting parameter set (C, D) is subsequently applied to the ³He nucleus without additional adjustments, with the Coulomb interaction included as described above.

The hopping amplitude is determined by the lattice spacing as $t = (\hbar c)^2/(2m_N a^2)$, yielding $t \approx 1.0$ MeV for $a = 4.5$fm and $m_N = 940$ MeV. The two-body low-energy constant is fixed by reproducing the experimental deuteron binding energy $EB_D = 2.224$ MeV, giving $C = -0.844$ MeV. The three-body low-energy constant is subsequently determined by fitting to the experimental triton binding energy $EB_T = 8.481$ MeV with $C$ held fixed, yielding $D =$

−5.000MeV. These parameter values are held fixed throughout all calculations presented in this work.

**3. Classical Exact Diagonalization**

We employ exact diagonalization (ED) to generate benchmark solutions for the nuclear models investigated in this study. Our calculations utilize a one-dimensional lattice model, where each site accommodates four spin–isospin single-particle orbitals,

$$\sigma \in \{p\uparrow, p\downarrow, n\uparrow, n\downarrow\}. \tag{5}$$

For a lattice with $N_s$ sites, this corresponds to a total of $4N_s$ fermionic modes. The Hamiltonian is represented in the occupation-number (Fock) basis constructed from these modes, and the Hilbert space is reduced by projecting onto subspaces with fixed particle numbers. All ED calculations were performed using custom Python code, employing the NumPy and SciPy libraries for matrix construction and numerical diagonalization.

The deuteron system is studied on a two-site lattice, while the triton and helium-3 systems are considered on three-site lattices. For the triton and helium-3 systems, the Hilbert space is further restricted to a fixed spin-projection sector, $S_z = +\frac{1}{2}$, which allows for a reduced and physically relevant model space. Because the Hamiltonian contains no explicit spin-dependent operators, the deuteron ground-state energy is independent of the chosen spin projection.

**Table 1.** Lattice configurations, particle-number constraints, spin-projection sectors (where applicable), and the resulting matrix dimensions used in the classical exact diagonalization (ED) calculations.

| System | $N_{sites}$ | $N_{particles}$ | $S_Z$ | ED matrix dimension |
|---|---|---|---|---|
| **Deuteron** | 2 | 2 | Not constrained | 16 |
| **Triton** | 3 | 3 | +1/2 | 36 |
| **Helium-3** | 3 | 3 | +1/2 | 36 |

The ED matrix dimension for the deuteron is obtained by restricting the Hilbert space to the physical one-proton, one-neutron sector, yielding C(4,1) × C(4,1) = 16 states. For the triton and helium-3 systems, the three-nucleon subspace with fixed spin projection S_z = +1/2 results in a matrix dimension of 36. This dimension is obtained by enumerating all three-particle configurations from the 12 available spin-isospin modes (3 sites × 4 modes) satisfying simultaneously the particle-number constraint (N = 3), the isospin composition (1p+2n for triton, 2p+1n for helium-3), and the spin-projection constraint (S_z = +1/2), yielding 36 physical states in each case.

The lattice EFT Hamiltonian $H_{\mathrm{EFT}}$, including the Coulomb interaction for the helium-3 system, is diagonalized numerically to obtain the ground-state energy $E_0^{\mathrm{ED}}$. The ED calculations employ the same Hamiltonian parameters (C, D) and the same model space as those used in the variational quantum simulations.

In addition to ground-state energies, selected low-lying eigenstates and dominant occupation-number configurations are analyzed. These results serve as classical reference data for benchmarking the variational quantum calculations.

## 4. Variational Quantum Eigensolver

We utilize the Variational Quantum Eigensolver (VQE) as a benchmark to evaluate the performance of variational quantum algorithms for few-nucleon systems governed by lattice effective field theory (EFT) Hamiltonians. For the VQE calculations, the second-quantized lattice EFT Hamiltonian is mapped onto a qubit operator via the Jordan–Wigner (JW) transformation [5,10]. The full qubit Hamiltonian is given by:

$$H_{\text{qubit}} = H_{\text{kin}}^{\text{JW}} + H_{\text{2B}}^{\text{JW}} + H_{\text{3B}}^{\text{JW}} + H_{\text{Coul}}^{\text{JW}}. \tag{6}$$

where the individual terms are:

$$H_{kin}^{JW} = \frac{1}{4Ma^2} \sum_{\langle \mathrm{i,j} \rangle} \sum_{\sigma} \left( \mathrm{X}_{\mathrm{q}_{\mathrm{i}\sigma}} \mathrm{Z}_{\mathrm{q}_{\mathrm{i}\sigma}+1} \cdots \mathrm{Z}_{\mathrm{q}_{\mathrm{j}\sigma}-1} \mathrm{X}_{\mathrm{q}_{\mathrm{j}\sigma}} + \mathrm{Y}_{\mathrm{q}_{\mathrm{i}\sigma}} \mathrm{Z}_{\mathrm{q}_{\mathrm{i}\sigma}+1} \cdots \mathrm{Z}_{\mathrm{q}_{\mathrm{j}\sigma}-1} \mathrm{Y}_{\mathrm{q}_{\mathrm{j}\sigma}} \right). \tag{7}$$

$$H_{2B}^{JW} = \frac{C}{8} \sum_{i} \sum_{\sigma \neq \sigma'} \left(1 - Z_{q_{i\sigma}}\right) \left(1 - Z_{q_{i\sigma'}}\right). \tag{8}$$

$$H_{3B}^{JW} = \frac{D}{48} \sum_{i} \sum_{\sigma \neq \sigma' \neq \sigma''} \left(1 - Z_{q_{i\sigma}}\right) \left(1 - Z_{q_{i\sigma'}}\right) \left(1 - Z_{q_{i\sigma''}}\right). \tag{9}$$

$$H_{\text{Coul}}^{\text{JW}} = \frac{1}{4} \sum_{i<j} V_{ij} \sum_{\sigma,\sigma' \in \{\uparrow p, \downarrow p\}} (1 - Z_{q_{i\sigma}})(1 - Z_{q_{j\sigma'}}). \tag{10}$$

Here, $q_{i\sigma}$ denotes the qubit index associated with spin-isospin mode $\sigma$ at lattice site $i$. The classical exact diagonalization (ED) calculations are performed directly in the fermionic Fock space without applying the JW transformation. The ED and VQE calculations thus operate on equivalent Hamiltonians expressed in different representations, and their results are compared to benchmark the performance of the variational algorithm.

### 4.1 Variational ansatz design

The variational circuits are designed to operate within subspaces of fixed particle number and, where applicable, fixed spin projection S_z, thereby restricting the optimization to physically admissible sectors of the Hilbert space. For each system, the variational circuit is initialized in a computational basis state informed by the dominant occupation-number configuration obtained from the classical ED calculation. The corresponding qubits are prepared in the $|1\rangle$ state via X-gates, while the remaining qubits are initialized to $|0\rangle$. This initialization strategy ensures that the variational optimization begins within a physically relevant region of the Hilbert space, thereby facilitating convergence.

For the deuteron system, the ansatz consists of two-qubit entangling blocks acting on selected qubit pairs within the physical sector. A representative block is shown in Figure 1. The full variational circuit operates on eight qubits, corresponding to two lattice sites with four spin-isospin modes each, and is constructed by applying iterative applications of this entangling block.

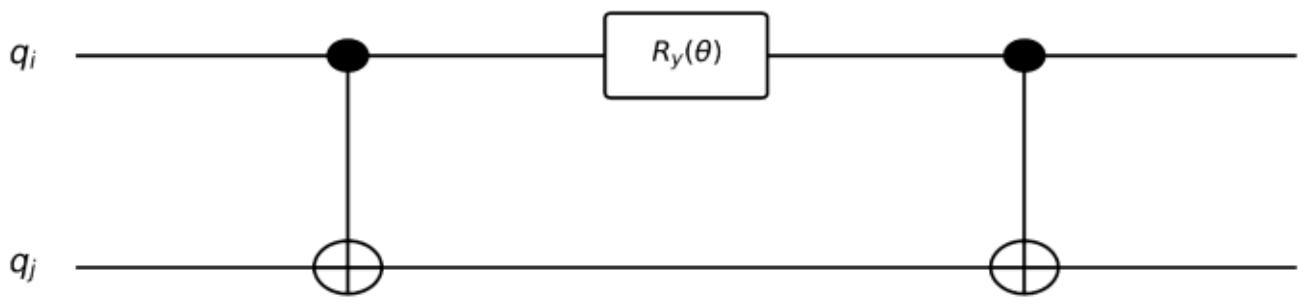

**Figure 1**. Two-qubit entangling block employed in the deuteron ansatz. This block comprises CNOT gates and a parameterized rotation R(θ), and functions as the elementary construct within the full 8-qubit (two sites × four modes) variational quantum circuit constrained to the physical two-particle sector.

The variational layer, shown schematically in Figure 2, encodes fermionic hopping via Givens rotations and same-site two-body correlations through $R_{ZZ}$ gates. The three-body interaction term $H_{3B}^{\mathrm{JW}}$, which in the qubit Hamiltonian contributes Z-type correlations among three modes within a single lattice site, is represented in the variational circuit using an effective $R_{ZZ}$-based approximation. A direct implementation using three-qubit diagonal phase gates was found to lead to optimization instabilities; the effective two-qubit representation was therefore adopted as a practical ansatz choice that captures the dominant on-site three-body correlations while maintaining a tractable optimization landscape.

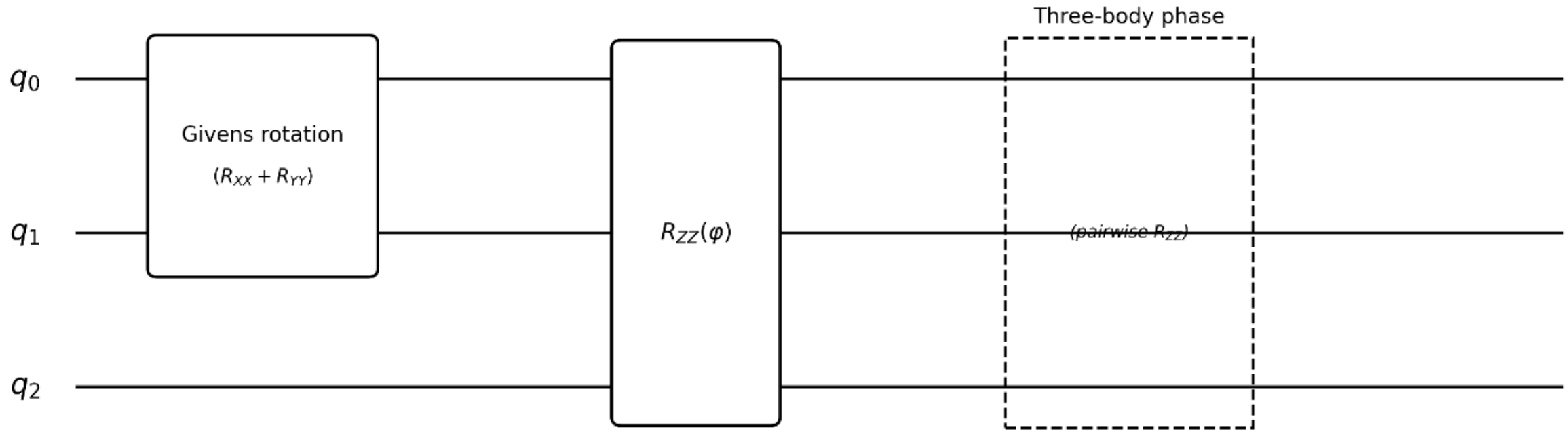

**Figure 2.** Schematic of a particle-number--conserving ansatz layer for three-nucleon systems. The full ansatz operates on 12 qubits (three sites × four modes), with qubits ordered as $q_{i,\sigma}$ where $\sigma \in \{p\uparrow, p\downarrow, n\uparrow, n\downarrow\}$ for each site $i$. Three representative qubits are shown for clarity. Fermionic hopping between adjacent sites is encoded via Givens rotations $(R_{XX} + R_{YY})$ acting on qubit pairs sharing the same spin-isospin mode $\sigma$. On-site two-body correlations between different modes at the same site are captured via $R_{ZZ}(\varphi)$ gates. Three-body interactions are represented via an effective $R_{ZZ}$-based phase operator acting within a single lattice site, shown in the dashed box.

In practice, the particle-number--conserving layer is applied to qubits associated with nearest-neighbor lattice sites, where Givens rotations act on pairs of qubits encoding the same spin-isospin mode $\sigma$ at adjacent sites (i, j), and to same-site qubit pairs where $R_{ZZ}$ gates encode on-site two-body correlations between different spin-isospin modes.

The number of variational parameters scales with the number of ansatz layers. Each layer comprises rotation angles from Givens rotations, two-body correlation gates, and an on-site phase operator that encodes three-body effects. Consequently, the circuit depth increases approximately linearly with layer count. The specific qubit counts and variational parameter values employed in these simulations are detailed in Table 2.

**Table 2.** Qubit counts and variational parameter numbers used in the VQE simulations for the deuteron, triton, and helium-3 systems. The number of variational parameters increases with the number of ansatz layers and corresponds to the increasing complexity of the nuclear systems considered. The three additional variational parameters in the helium-3 ansatz relative to the triton arise from proton–proton same-site RZZ correlation gates, introduced to capture the on-site Coulomb repulsion between the protons and are absent in the triton ansatz.

| System | Qubits | Variational parameters per layer | Layers | Total parameters |
|---|---|---|---|---|
| **Deuteron** | 8 | 8 | 2 | 16 |
| **Triton** | 12 | 18 | 3 | 54 |
| **Helium-3** | 12 | 21 | 3 | 63 |

The increase in parameter count from the deuteron to three-nucleon systems arises from the expansion of the Hilbert space and the necessity to incorporate additional correlation structures within the variational circuit. The modestly higher parameter count in the helium-3 ansatz is attributable to the additional correlation structures introduced by the repulsive Coulomb interaction among protons.

The entanglement structure of the variational circuit is designed to reflect the locality of the lattice EFT Hamiltonian. Two-qubit gates are predominantly applied between qubits associated with neighboring lattice sites and between spin–isospin modes located on the same lattice site. This design decision ensures that the variational circuit encodes the dominant physical correlations, thereby circumventing the introduction of spurious long-range entanglement, which would increase circuit depth without corresponding physical justification.

The ansatz therefore constitutes a subclass of physics-inspired variational circuits, whose gate structure is derived from the dominant dynamical processes of the underlying Hamiltonian. Specifically, fermionic hopping terms are implemented via Givens rotations, while on-site interaction terms are encoded using controlled-phase gates applied to qubits residing on the same lattice site. In contrast to hardware-efficient ansätze, which apply generic parameterized rotations without regard to the underlying symmetries, the triton and helium-3 constructions enforce particle-number conservation by design, through Givens rotations for fermionic hopping and diagonal RZZ gates for on-site interactions — both of which preserve particle number by construction. For the deuteron system, physical sector confinement is ensured through initialization in the dominant occupation-number configuration identified by classical ED, with X-gates preparing the relevant qubits in the $|1\rangle$ state prior to variational optimization.

### 4.2 Optimization and eigenstate validation

We optimize the VQE using the derivative-free classical optimizer COBYLA, on a classical statevector simulator. We employ a layer-by-layer strategy, initializing deeper circuits with parameters optimized at shallower depths. This approach enhances numerical stability and facilitates convergence when increasing circuit depth. The convergence behavior for the triton system is shown in Figure 3.

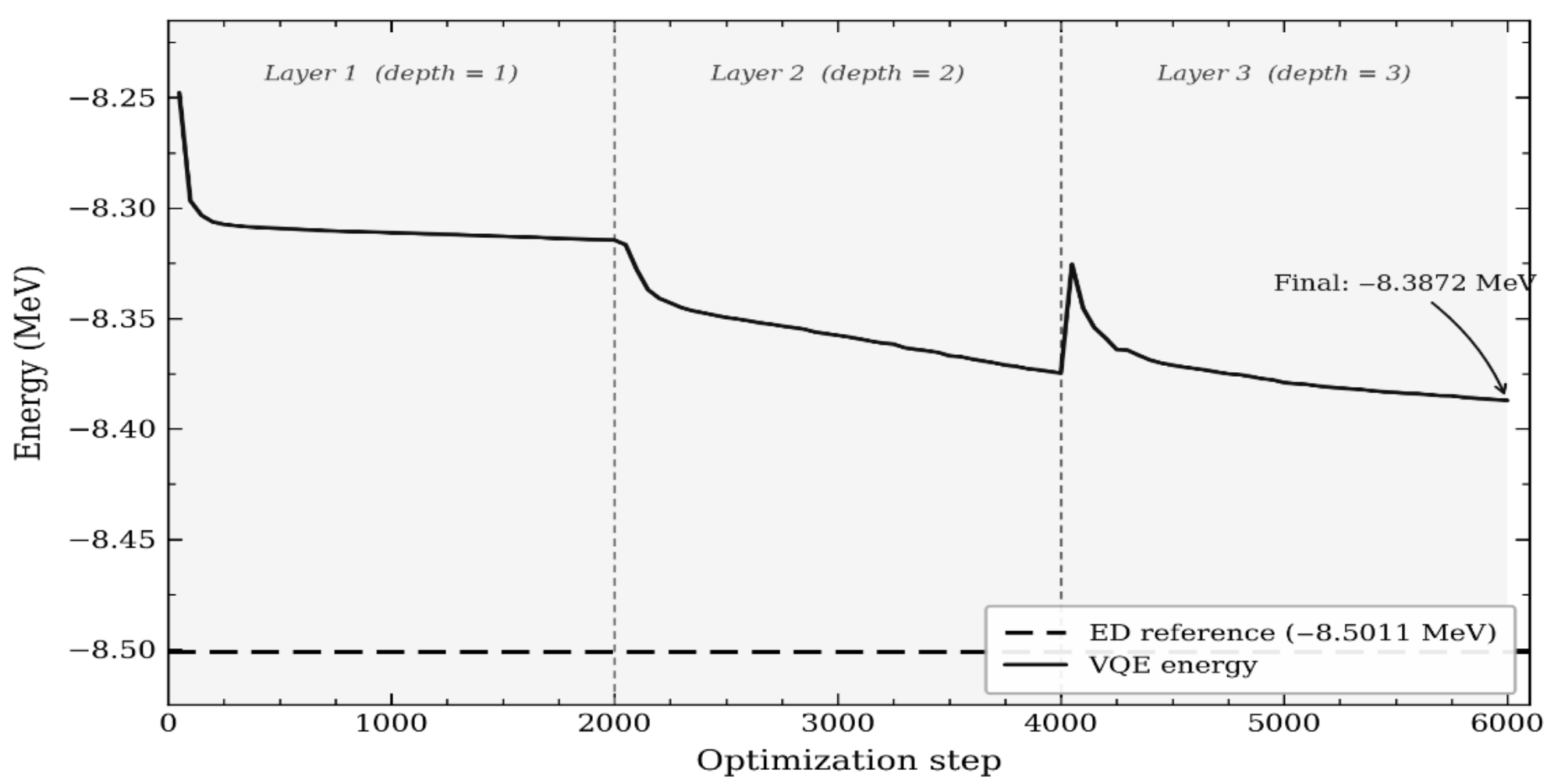

**Figure 3.** The variational energy at the start of Layer 1 is already close to the ED reference, reflecting the physically informed initialization of the circuit in the dominant ED configuration. Variational parameters are initialized to small random values drawn uniformly from $[-0.2, 0.2]$ for the first layer and $[-0.1, 0.1]$ for subsequent layers, with a fixed random seed (seed = 42) to ensure reproducibility.

Energy minimization alone does not ensure convergence to an exact eigenstate. To assess the quality of the variational wavefunction, we compute the Hamiltonian variance

$$\sigma^2(H) = \langle H^2 \rangle - \langle H \rangle^2 \tag{11}$$

This variance vanishes for an exact eigenstate. The variance is therefore used as a diagnostic quantity to assess the closeness of the variational state to an eigenstate and to quantify the admixture of excited-state components [18,19].

The reported VQE observables include the optimized ground-state energy, the corresponding energy variance, and the dominant configurations of the variational state. These results are compared with the classical ED calculations in the subsequent section.

### 4.3 Noisy VQE Simulations

To investigate the robustness of the variational algorithm under realistic hardware imperfections, additional simulations were performed using a noisy quantum circuit model. These simulations were designed to emulate the effects of gate errors and finite sampling statistics that arise on current noisy intermediate-scale quantum (NISQ) devices.

The noisy simulations were implemented using the Qiskit Aer simulator with a depolarizing noise model applied to both single-qubit and two-qubit gates. Experimental superconducting-qubit platforms have reported single-qubit gate fidelities above 99.9% and two-qubit fidelities around 99.4% [20]. In line with these scales, we adopt depolarizing probabilities $p_1 = 3 \times 10^{-4}$ and $p_2 = 3 \times 10^{-3}$ in the present simulations.

Expectation values of the Hamiltonian were evaluated using a shot-based estimator. During the variational optimization, 2048 measurement shots were used to estimate the energy expectation value, while 4096 shots were used for the final evaluation of the optimized circuit. This approach balances computational efficiency during optimization with improved statistical accuracy in the final reported results.

To mitigate the exponential growth of the variational parameter space in the presence of shot noise, we employed a layer-by-layer training strategy. In this approach, parameters optimized for a given circuit depth are frozen while additional parameters for the next variational layer are sequentially introduced and optimized. This strategy mitigates barren plateaus, stabilizes the optimization landscape, and reduces the computational cost incurred by noisy function evaluations.

Because increasing circuit depth under shot noise rapidly expands the variational parameter space, exacerbates susceptibility to barren plateaus, and leads to excessively long convergence times, the noisy simulations presented in this work are restricted to the first variational layer. This constitutes a deliberate methodological choice to isolate the effect of gate-level noise on energy estimation while maintaining a tractable optimization problem.

While the noiseless statevector simulation reproduces the exact diagonalization (ED) reference energy with high fidelity, the noisy simulation yields a significantly larger deviation attributable to depolarizing gate errors and shot-based measurement statistics. Despite these errors, the variational optimization remains robust, yielding an energy value that falls within the physically plausible regime.

## 5. Results

This section presents a comparison of ground-state energies obtained from classical exact diagonalization (ED) and variational quantum eigensolver (VQE) calculations for the deuteron, triton, and helium-3 systems within the lattice pionless EFT framework. The ED and VQE energies, accompanied by the absolute energy differences and Hamiltonian variances, are summarized in Table 3. These results are visualized in Figure 4.

**Table 3.** Ground-state energies obtained from classical exact diagonalization (ED) and variational quantum eigensolver (VQE) calculations for the deuteron, triton, and helium-3 systems. The absolute energy difference $|\Delta E| = |E_{\mathrm{VQE}} - E_{\mathrm{ED}}|$ and the Hamiltonian variance $\mathrm{Var}(H) = \langle H^2 \rangle - \langle H \rangle^2$ are reported as diagnostics of variational accuracy and proximity to an eigenstate. The noisy VQE result is reported only for the triton system, as it serves as a representative three-body benchmark for assessing the impact of realistic NISQ-era noise; extending noisy simulations to helium-3 is left for future work.

| System | E_ED (MeV) | E_VQE (MeV) | E_VQE noisy (MeV) | \|ΔE\| (MeV) | Var(H) (MeV$^2$) |
|---|---|---|---|---|---|
| **Deuteron** | −2.222 | −2.222 | - | 0.000 | 0.000 |
| **Triton** | −8.501 | −8.387 | -8.032 | 0.114 | 0.410 |
| **Helium-3** | −7.898 | −7.765 | - | 0.133 | 0.481 |

Hamiltonian variance is not reported for the noisy VQE result. In shot-based simulations, the estimator conflates intrinsic quantum variance with statistical sampling noise arising from finite measurement statistics, rendering the variance diagnostic uninterpretable as a pure eigenstate quality measure. The noisy VQE energy is therefore reported without an associated variance.

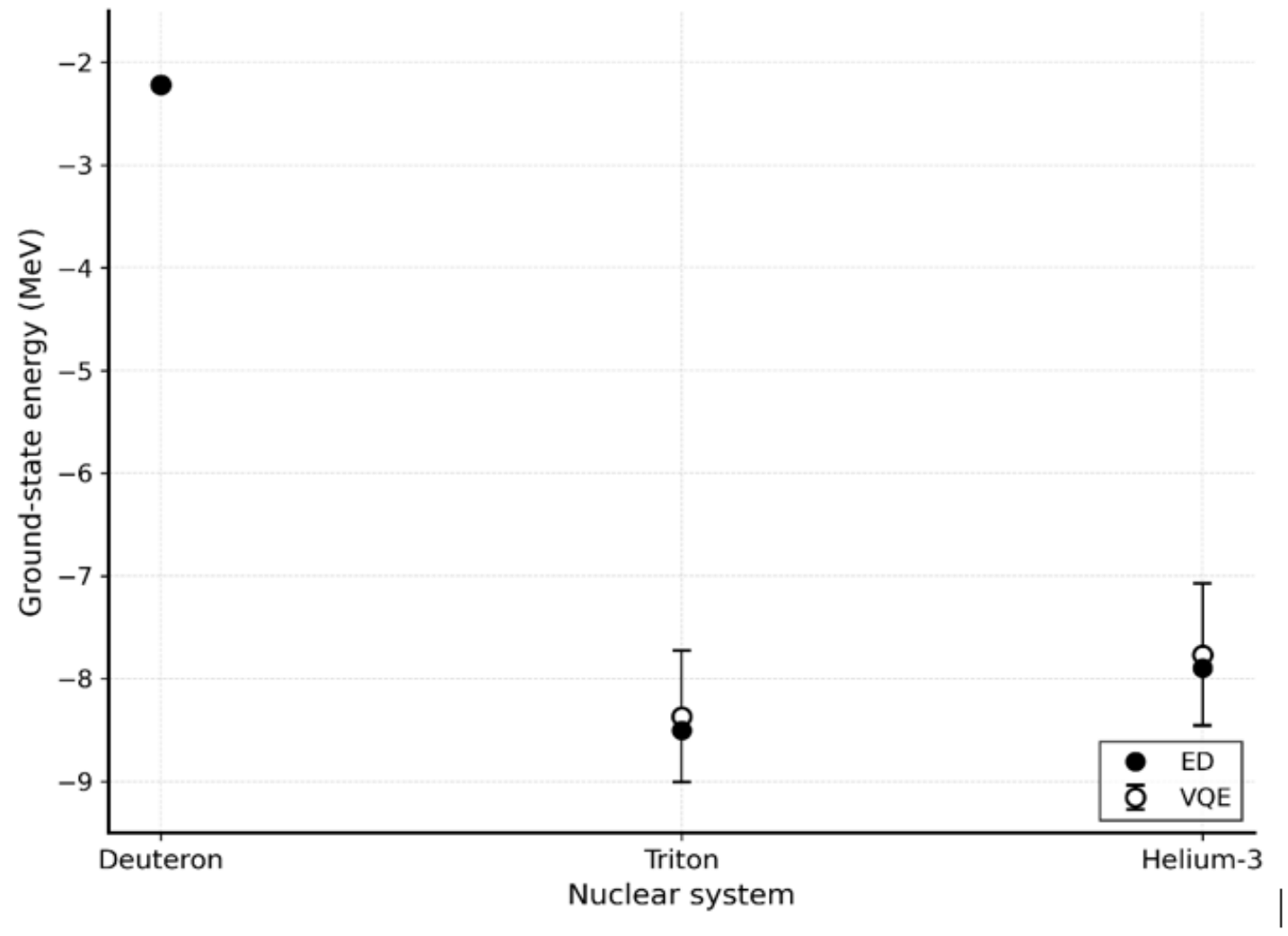

**Figure 4.** Ground-state energies obtained from classical exact diagonalization (ED, filled circles) and noiseless statevector simulations of the Variational Quantum Eigensolver (VQE) (open circles) calculations for the deuteron, triton, and helium-3 systems. Energies are given in MeV. Error bars on the VQE results quantify the variability in the variational energy during the final optimization steps. The close agreement between ED and VQE values across all three systems validates the accuracy of the variational approach within the lattice pionless EFT framework.

### 5.1 Deuteron

The deuteron system serves as a minimal validation case for the VQE workflow due to its small Hilbert space and the availability of an exact classical reference. Using the lattice Hamiltonian described in Section 2, the ED ground-state energy is found to be $E_{\mathrm{ED}} = -2.222$ MeV, while the corresponding VQE calculation yields $E_{\mathrm{VQE}} = -2.222$ MeV (Table 3).

The resulting energy difference is negligible within numerical precision ($\mid \Delta E \mid = 0.000$ MeV), and the associated Hamiltonian variance, $\mathrm{Var}(H) = 0.00\ MeV^2$, is consistent with zero within numerical precision, indicating that the variational state effectively coincides with an eigenstate of the Hamiltonian. Throughout the optimization, the VQE wavefunction is found to remain confined to the physical two-particle subspace, with no significant indication of leakage.

The deuteron system thus serves as a controlled reference for validating the Hamiltonian implementation, qubit mapping, and optimization procedure, and establishes a baseline for analyzing more complex nuclear systems.

### 5.2 Triton

The triton system constitutes a more demanding benchmark due to the presence of genuine three-body correlations. The ED calculation yields a ground-state energy of $E_{\mathrm{ED}} = -8.501$MeV, while the VQE approach produces $E_{\mathrm{VQE}} = -8.387$ MeV, resulting in an energy difference of $\mid \Delta E \mid = 0.114$ MeV (Table 3).

The Hamiltonian variance of the VQE state is $\mathrm{Var}(H) = 0.41\ MeV^2$, which is finite and larger than in the deuteron case. Particle-number conservation is maintained throughout the optimization, and the variational state remains confined to the physical three-nucleon subspace (one proton and two neutrons).

The deviation from the exact diagonalization (ED) reference underscores the heightened sensitivity of the three-body system to the precision of variational parameter optimization relative to the two-body deuteron system. The non-zero variance signifies that the optimized variational wavefunction successfully approximates the dominant low-energy eigenstate characteristics, yet retains measurable admixtures from higher-energy excitations.

In addition to the noise-free statevector simulation, a noisy VQE calculation was conducted for the triton system using a depolarizing noise model to simulate realistic NISQ hardware conditions. The resulting noisy VQE ground-state energy was −8.032 MeV. This value deviates from the noiseless VQE result (−8.387 MeV) and the exact diagonalization reference energy (−8.501 MeV). A comparison of the noise-free and noisy results is presented in Figure 5.

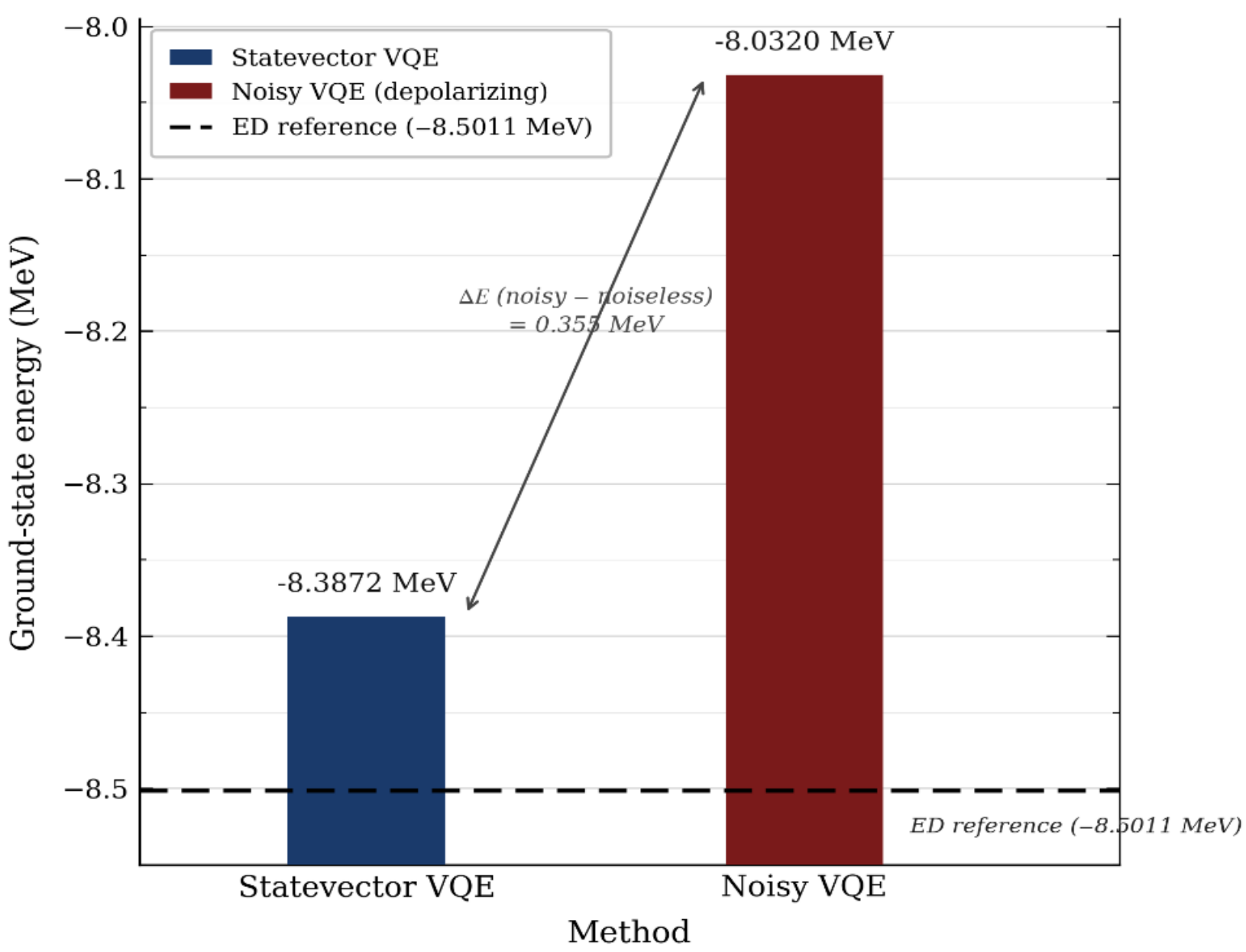

**Figure 5.** Comparison of ground-state energies obtained from noiseless statevector VQE and noisy VQE simulations for the triton system, illustrating the impact of depolarizing noise and finite sampling (shot noise) on the VQE energy estimate. The horizontal dashed line indicates the exact diagonalization (ED) reference energy.

Notably, the noisy simulation was performed using only the first ansatz layer, as extending to deeper circuits under shot-based estimation led to excessively long convergence times and erratic optimization dynamics. Consequently, the Hamiltonian variance is not reported for the noisy simulation. As detailed in Section 4.3, shot-based estimators cannot disentangle intrinsic quantum variance from finite-sampling noise, rendering the computed quantity an unreliable eigenstate diagnostic.

### 5.3 Helium-3

The helium-3 system exhibits additional complexity due to the Coulomb repulsion between the two protons. The ED ground-state energy is $E_{\mathrm{ED}} = -7.898$ MeV, while the corresponding VQE calculation yields $E_{\mathrm{VQE}} = -7.765$ MeV, resulting in an energy difference of $|\Delta E| = 0.133$ MeV (Table 3).

The Hamiltonian variance is $\mathrm{Var}(H) = 0.48\ MeV^2$, which is slightly larger than in the triton case. This behavior is consistent with the increased complexity of the helium-3 Hamiltonian, which includes both three-body interactions and Coulomb repulsion. Relative to the triton, the helium-3 ground-state energy is systematically shifted upward, in line with the repulsive Coulomb contribution included in the model.

As demonstrated for the triton system, particle-number conservation is maintained throughout the VQE optimization, and the variational wavefunction remains constrained to the physical three-nucleon subspace (two protons and one neutron).

## 6. Discussion

This work establishes a systematic benchmark to evaluate variational quantum algorithms across few-body nuclear systems of escalating complexity within a unified lattice EFT framework.

For the few-body systems considered here, classical simulation of exact diagonalization remains computationally tractable. Consequently, this work should be viewed as an algorithmic benchmark rather than a demonstration of quantum advantage.

For the deuteron system, the exact concordance between ED and VQE energies and the negligible Hamiltonian variance demonstrate that the variational circuit fully spans the relevant two-particle Hilbert-space sector. This result confirms the internal consistency of the computational pipeline, including the fermion-to-qubit mapping, Hamiltonian construction, and classical optimization procedure. The deuteron calculation therefore serves as an essential validation step for the methodology before extending the framework to more complex nuclear systems.

For the triton and helium-3 systems, the variational results exhibit modest yet finite deviations from the ED reference energies (approximately 0.13 MeV) along with non-vanishing Hamiltonian variances. These deviations align with the heightened entanglement structure of three-body systems and the constrained expressibility inherent to shallow variational quantum circuits. The ansatz effectively captures the dominant low-energy configurations yet fails to fully account for all many-body correlations arising from the complex interplay between hopping processes and on-site interactions across multiple lattice sites.

Importantly, particle-number conservation is enforced by construction in the ansatz, ensuring that the observed deviations are attributable to variational expressibility rather than leakage into unphysical sectors. This constraint is particularly relevant for nuclear simulations, where particle-number conservation is a fundamental symmetry of the underlying Hamiltonian.

The low-energy constants are calibrated in a sequential hierarchy: the two-body coupling $C$ is first determined by reproducing the experimental deuteron binding energy, after which the three-body coupling $D$ is fixed from the triton binding energy. The resulting parameter set (C, D) is then applied without modification to the helium-3 nucleus. The resulting ED and VQE energies reproduce the expected qualitative physics: the Coulomb repulsion between the two protons shifts the helium-3 binding energy upward relative to the triton. This behavior confirms that the quantum simulation captures physically meaningful features of the underlying effective interaction rather than merely fitting individual systems.

The noisy VQE simulations further highlight the limitations imposed by NISQ-era hardware. The depolarizing noise model used here reflects typical error rates of contemporary superconducting quantum processors [20]. Under these conditions, the triton energy shifts by approximately 0.355 MeV relative to the noiseless result (from −8.387 MeV to −8.032 MeV, corresponding to a relative deviation of approximately 4.2% with respect to the ED reference), demonstrating that accumulated gate errors can significantly impact energy estimation even for moderate circuit depths. Nevertheless, the optimization procedure remains stable and does not diverge from the physically relevant energy scale. The layer-by-layer training strategy was adopted to stabilize the variational optimization landscape and reduce the computational overhead associated with noisy circuit evaluations. Alternative training protocols, such as simultaneous full-circuit optimization or adaptive ansatz construction methods such as ADAPT-VQE [21], as well as error mitigation techniques such as zero-noise extrapolation (ZNE) [22] and probabilistic error cancellation (PEC) [23], represent natural directions for future investigation.

From a resource standpoint, the present simulations operate within the regime accessible to near-term quantum hardware. However, extending the framework to larger nuclei or higher-dimensional lattices would necessitate substantially larger qubit registers and deeper circuits, thereby significantly exacerbating noise effects. As a concrete illustration of resource scaling, each additional lattice site contributes four qubits to the register, so that an $N_s$-site system requires $4N_s$ qubits in total. The number of Pauli operator terms arising from two-body interactions grows as $\mathcal{O}(N_s^2)$, while three-body terms contribute $\mathcal{O}(N_s^3)$ additional operators. For the present systems ($N_s = 2,3$), these counts remain tractable; however, already at $N_s = 6$–8, the qubit requirement reaches 24–32 and the Pauli term count increases by roughly an order of magnitude, placing such simulations at the boundary of near-term hardware capability. Consequently, the development of improved ansätze, symmetry-preserving circuit architectures, and advanced error-mitigation protocols will be imperative for scaling quantum nuclear simulations beyond few-body benchmark systems.

## 7. Conclusion

This work presents a systematic benchmark study of the Variational Quantum Eigensolver applied to light nuclei within a lattice formulation of pionless effective field theory (EFT). By combining classical exact diagonalization with variational quantum algorithms, we establish a systematic benchmarking framework for evaluating the accuracy and robustness of quantum algorithms for few-body nuclear problems.

Across all three systems, the VQE results are in substantial agreement with the classical exact diagonalization reference energies. The deuteron benchmark confirms the internal consistency of the full computational pipeline, while the modest yet non-zero deviations observed for the triton and helium-3 systems (≈0.13 MeV) are consistent with the restricted expressibility of shallow variational circuits acting in larger Hilbert spaces.

The hierarchical calibration of EFT parameters — fixing the two-body coupling from the deuteron and the three-body coupling from the triton — permits a consistent prediction of the helium-3 binding energy including Coulomb interactions. This result illustrates that the variational quantum simulation captures meaningful physical features of the underlying effective interaction as opposed to simply reproducing isolated benchmark values.

Noisy VQE simulations indicate that depolarizing gate errors introduce a relative error of approximately 4.2% for the three-body system compared to the exact diagonalization (ED) reference. Despite this error, the variational optimization remains stable and converges within the physically relevant energy scale. This stability suggests that structured ansatz designs and layer-wise training protocols may be compatible with noisy circuit execution on near-term quantum devices.

Overall, the present study demonstrates that variational quantum algorithms can reproduce the ground-state energies of light nuclear systems with high accuracy within the framework of lattice effective field theory (EFT). While the present calculations remain within a regime accessible to classical diagonalization, they provide a controlled environment for testing quantum algorithms and circuit designs for nuclear many-body problems.

This work therefore provides a transparent and reproducible benchmark for variational quantum simulations of few-body nuclear systems.

## Statements and Declarations

### Author Contributions

P. Çifci conceived the study, developed the theoretical framework, implemented all numerical codes, performed the VQE and exact diagonalization calculations, analyzed the results, and wrote the manuscript. S. Akkoyun supervised the work and critically reviewed the manuscript. All authors read and approved the final manuscript.

### Competing Interests

The authors declare no competing interests.

### Funding

This research received no external funding.

### Data Availability Statement

The numerical codes used in this study are available from the corresponding author upon reasonable request.